# Open-source software for studying neural codes

*Robin A. A. Ince*

*Max-Planck Institute for Biological Cybernetics, Tübingen, Germany*
*robin.ince@tuebingen.mpg.de*

## Introduction

In recent years, neuroscience has become an increasingly computational discipline. This is a natural consequence of the complexity of the systems under study combined with the constant improvement and development of novel experimental techniques, which allow for increasingly detailed observations. Neurophysiologists can now record simultaneous neural activity, at temporal resolutions of tens of kHz, from tens to hundreds of intracranial electrodes (Csicsvari et al., 2003). From each electrode, both action potentials of individual neurons (reflecting the output of a cortical site, see chapters 3,5,6) and local field potentials (LFPs; reflecting both population synaptic potentials and other types of slow activity, see chapter 4) can be extracted. Moreover, electrophysiological recordings can now be accompanied by simultaneous measurements of other brain signals, such as those recorded with optical imaging, electroencephalography (EEG, see chapter 23) or functional magnetic resonance imaging (fMRI, see chapter 25). Such developments make managing the collection, storage, pre-processing and analysis of such data a significant computational challenge. Further, increasingly detailed large-scale modelling produces sizable quantities of synthetic data that must also be carefully analysed to give insight into the behaviour of the models and to provide meaningful comparisons to experiments.

Analysis techniques must be advanced to keep pace with these developments, both in terms of dealing with the quantities of data now available, but also by developing more sensitive and powerful methods to derive the maximum benefit from the collected data. Indeed, development of novel techniques can breathe new life into archived data, allowing new questions to be addressed with results from old experiments. There is a strong argument that the development of tools for analysing neurophysiological data and for performing quantitative investigation of neural coding would be helped by standardization and public and transparent availability of the software implanting these analysis tools (Ince et al., 2012), together with sharing of experimental data (Ascoli, 2006; Teeters et al., 2008), as well as by the standardization of experimental and modelling neuroscience procedures (Nordlie et al., 2009).

In this chapter we first outline some of the popular computing environments used for analysing neural data, followed by a brief discussion of 'software carpentry', basic tools and skills from software engineering that can be of great use to computational scientists. We then introduce the concept of open-source software and explain some of its potential benefits for the academic community before giving a brief directory of some freely available open source software packages that address various aspects of the study of neural codes. While there are many commercial offerings that provide similar functionality, we concentrate here on open source packages, which in addition to being available free of charge, also have the source code available for study and modification. While we have made every effort to provide a comprehensive list of the most widely used and actively developed general packages, we apologise if we have inadvertently omitted any important contributions. We would also like to point out the INCF Software Center[1], a broader online database of neuroscience related software.

## Popular computing environments for neuroscience

In our experience, by far the most commonly used platform in computational neuroscience is MATLAB[2]. MATLAB is a numerical computing environment and programming language, which provides a flexible and interactive way to manipulate and visualise data, develop algorithms and perform analyses. Some advantages of MATLAB are its polished and tightly integrated development environment including debugging and profiling capabilities, together with a large collection of additional toolboxes providing specialised functionality. While this is a commercial package, and so might seem out of place in a chapter focussing on open source software, it's ubiquity and wide availability through campus licenses mean that much of the software considered here is built with it. MATLAB is a commercial product but discounted student licenses are available. There are also open source clones, such as Octave[3], which strive to maintain compatibility with MATLAB allowing access to analysis tools for those without a MATLAB license.

Another environment, which is growing in popularity within the neuroscience community (Koetter et al., 2009), is the programming language Python[4]. In recent years excellent numerical and scientific components (such as NumPy, SciPy and related projects[5]) have been developed which now make the language suitable for scientific computation, interactive data analysis and visualisation. Some advantages are that Python is a well designed, general purpose, object oriented language which is easy to learn and use and which has, in addition to the numerical and scientific computing tools, a large range of libraries freely available for general computing tasks.

Finally, R[6] is a language and environment for statistical computing and graphics. Some advantages of R include its flexible data types and a comprehensive archive of an extremely wide range of statistical techniques. Both Python and R are open-source and freely available.

---

[1] http://software.incf.org/
[2] http://www.mathworks.de/
[3] http://www.gnu.org/software/octave/
[4] http://www.python.org/
[5] http://www.scipy.org/   http://www.scipy.org/Topical_Software
[6] http://www.r-project.org/

## Software Carpentry

Over the past 50 or so years, computer programming or software engineering has grown and matured as a profession. Many tools have been developed to help with the difficult problems of managing software projects as the complexity and amount of code and number of contributors grows. While these tools and methodologies are now well established among the professional software development community, the scientific community, consisting primarily of self-taught programmers, has understandably been slower to adopt them. However, investing a small amount of time to become familiar with some of these tools and techniques can lead to great increases in productivity (Wilson, 2006). Practises such as using version control software, systematic automated testing, scripting to automate repetitive tasks as well studying program design and techniques such as object-oriented development can all yield dramatic benefits. The Software Carpentry website[7] is an excellent resource providing free courses on these and other topics.

## Open-Source for Academic Software

Open-source is a software development method in which the source code is made available under a license that allows free distribution and the creation of derived works[8]. The primary advantages of open source software for an end user are its availability without cost, the availability of the source code, the right to modify the software and the right to redistribute such modifications.

These advantages have been recognised in the commercial sector, as evidenced by the growing popularity of open source software, for example the Linux operating system. However, in the academic setting the benefits of open-source are even more compelling. The importance of publically releasing analysis codes is beginning to get widespread attention (Peng, 2009; Barnes, 2010; Ince et al., 2012). The practice of distributing analysis software has significant benefits for the neuroscience community as a whole, both for experimentalists who would have a wider range of analysis techniques readily available to apply to their data, and for theoreticians, who would receive a wider audience for their techniques, providing greater feedback and testing as well as potentially allowing them to be applied in new ways and in different areas from those originally intended. It is also important, together with the sharing of published data sets, for the important scientific principle of reproducibility.

The availability of the source code, together with the rights to modify and distribute, provide additional important benefits; the code can be a valuable learning tool to improve understanding and aid with verification of the methods, and it also allows for customisation of the algorithms so that they might be extended to account for some new data type, unforeseen special case or to incorporate other improvements. Although it can take some extra effort to clean up and document code for public release, it is invariably a useful step, both for the authors themselves should they return to the code after some time and also for new members to the group. To address the practicalities of public distribution, it is now easier than ever to create a website, share files online, or set up a public code repository[9].

---

[7] http://software-carpentry.org/
[8] http://www.opensource.org/

As well as making the code available, to provide the additional rights the copyright holders release the code under a certain license, by adding the appropriate license text to each source file or by clearly including the license in the distribution of files. There are now many open-source compatible licenses, but here we describe briefly just the two most commonly used, the GNU General Public License (GPL) and the BSD License. These take two different approaches. The BSD license is the most permissive, allowing unlimited redistribution for any purpose as long as the copyright notices are maintained. In contrast, the GPL restricts in some way the use of derived works – they must also be freely distributed themselves under the terms of the GPL. There are passionate proponents of both sides, BSD supporters feel the 'viral' nature of the GPL is too strong and don't want to restrict use of their software in any way, while GPL users prefer to ensure their software remains 'free' and prevent users redistributing modified versions without source code and not 'giving back' their improvements to the community. However, much research software is currently released under custom non-standard licenses often including specific attribution or citation clauses, which can be incompatible with the common open-source licenses described above. They are also likely to offer less legal protection, since the common licences and now well established, widely understood and legally tested. We would urge researchers considering a code release to use one of the tried and tested standard licenses, together with a clear request in the documentation for a specific citation if the results are used in a published article.

## Software for Spike Sorting

Spike sorting is discussed in more detail in chapter 5.

### WaveClus

*http://www2.le.ac.uk/departments/engineering/research/bioengineering/neuroengineering-lab/spike-sorting*

WaveClus (Quian Quiroga et al., 2004) is a MATLAB package implementing a fast and unsupervised algorithm for spike detection and sorting. Features are selected as the wavelet coefficients whose distributions are furthest from a normal distribution and these features are clustered with super-paramagnetic clustering. It has both an interactive graphical user interface and a script based batch mode.

### MClust

*http://redishlab.neuroscience.umn.edu/MClust/MClust.html*

---

[9] See for example http://sourceforge.net/ http://code.google.com/ http://github.com/

MClust is a MATLAB toolbox with a graphical user interface which enables a user to perform automated and manual clustering on single-electrode, stereotrode and tetrode recordings. It allows selection of a range of features, as well as user defined features, manual cluster specification as well as integration with other clustering tools (for example KlusterKwik, see below). In addition to the graphical user interface it also has a script based batch mode.

**OSort**
*http://www.urut.ch/new/serendipity/index.php?/pages/osort.html*
OSort (Rutishauser et al., 2006) is a MATLAB package which implements a template based, unsupervised, online spike sorting algorithm.

**Spyke**
*http://www.swindale.ecc.ubc.ca/spyke*
Spyke (Spacek et al., 2008) is a Python package for visualization, navigation and spike sorting of extracellular waveform data.

**KlusterKwik**
*http://klustakwik.sourceforge.net/*
KlusterKwik (Harris et al., 2000) is a standalone program with a command line interface that implements an efficient algorithm for clustering with a Gaussian mixture model.

**SAC**
*http://portal.bm.technion.ac.il/Labs/niel/Pages/Software.aspx*
SAC (Shoham et al., 2003) is a MATLAB package with a graphical user interface implementing a mixture of t-distributions algorithm for clustering principle components of spike waveforms.

**EToS**
*http://etos.sourceforge.net/*
Efficient Technology of Spike-sorting (EToS) (Takekawa et al., 2010; Takekawa and Fukai, 2012) is a suite of command line programs for UNIX-like operating systems which implements filtering, spike detection, wavelet based feature extraction and classification via robust variational Bayes for a finite mixture of t-distributions. The code is parallelised to allow full use of multi-core workstations.

**SpikeOMatic**
*http://www.biomedicale.univ-paris5.fr/physcerv/C_Pouzat/newSOM/newSOMtutorial/newSOMtutorial.html*
SpikeOMatic is an R package for spike sorting which includes function for spike detection, feature extraction as well as many classical clustering algorithms.

## Software for Analysis of Spike Trains and Electrophysiology
**FieldTrip**
*http://fieldtrip.fcdonders.nl/*
FieldTrip (Oostenveld et al., 2011) is a MATLAB toolbox, primarily for the analysis of MEG and EEG data, supporting functions such as pre-processing, time-frequency analysis, source reconstruction, event-related analyses, statistical analysis and visualisation. It also includes support for real-time, online acquisition and analysis and some spike train analysis functionality.

**Chronux**
*http://chronux.org/*
Chronux (Mitra and Bokil, 2008) is a MATLAB toolbox providing many functions for the analysis of neural data, including preprocessing and filtering functions, spectral analysis and visualisation tools for point process and continuous signals.

**Spike Train Analysis Toolkit**
*http://neuroanalysis.org/neuroanalysis/goto.do?page=.repository.toolkit_home*
The Spike Train Analysis Toolkit (Goldberg et al., 2009) is a comprehensive MATLAB package for the information theoretic analysis of spike train data.

**FIND toolbox**
*http://find.bccn.uni-freiburg.de/*
The FIND toolbox (Meier et al., 2008) is a MATLAB toolbox providing functions for analysis of multiple-neuron recordings and network simulations.

**sigTOOL**
*http://sourceforge.net/projects/sigtool/*
sigTOOL (Lidierth, 2009) provides a user-extendable signal analysis environment for processing electrophysiological data within MATLAB. Neuron spike-train, as well as time and frequency analyses are built in.

**infoToolbox**
*http://www.infotoolbox.org/*
infoToolbox (Magri et al., 2009) is a MATLAB toolbox implementing a range of bias corrected estimates of information theoretic quantities, that can be applied to both continuous signals as well as spike train recordings.

**pyEntropy**
*http://code.google.com/p/pyentropy/*
pyEntropy (Ince et al., 2009) is a Python package implementing a range of bias corrected estimates of information theoretic quantities for discrete systems (spike trains or binned continuous signals) that also includes an algorithm for calculating maximum entropy distributions under marginal constraints, which can be used to quantitatively study high order interactions.

**OpenElectrophy**
*http://neuralensemble.org/trac/OpenElectrophy*
OpenElectrophy (Garcia and Fourcaud-Trocmé, 2009) is a software framework written in Python to facilitate data management and provide functionality and interfaces for analysis and visualisation of data. It contains analysis tools for both spike trains and LFP recordings. See also http://neuralensemble.org/ which hosts this project as well as other related neural simulation and modelling software.

**CSDPlotter**
*http://arken.umb.no/~klaspe/iCSD.php*

CSDPlotter (Pettersen et al., 2006) is a MATLAB toolbox implementing the inverse current source density method (iCSD) with a graphical user interface.

**iCSD 2D**
*http://www.neuroinf.pl/Members/szleski/csd2d/toolbox*
iCSD 2D (Łęski et al., 2011) is a MATLAB toolbox implementing the iCSD method in two dimensions.

**Neuropy**
*http://www.swindale.ecc.ubc.ca/neuropy*
Neuropy (Spacek et al., 2008) is a Python package providing functions for analysis of spike trains.

**Relacs**
*http://relacs.sourceforge.net/index.html*
Relacs (Benda et al., 2007) is a software platform for closed-loop data acquisition, online analysis and stimulus generation, specifically designed for electrophysiological recordings. Relacs is written in C++ and runs on UNIX-like systems.

**Neuroscope**
*http://neuroscope.sourceforge.net/*
Neuroscope (Hazan et al., 2006) is an advanced viewer for electrophysiological and behavioural data: it can display LFP, EEG, neuronal spikes, behavioural events as well as the position of the animal in the environment. Neuroscope is a stand-alone program for UNIX-like platforms.

**STAR**
http://sites.google.com/site/spiketrainanalysiswithr/
STAR (Pouzat and Chaffiol, 2009) is an R package to analyse spike trains from single or multiple simultaneously recorded neurons. It provides tools to visualise spike trains and fit, test and compare models of discharge applied to actual data.

## Software for Analysis of EEG
**EEGLAB**
http://sccn.ucsd.edu/eeglab/
EEGLAB (Delorme and Makeig, 2004) is an interactive MATLAB toolbox for processing continuous and event-related EEG, MEG and other electrophysiological data incorporating independent component analysis (ICA), time/frequency analysis, artefact rejection, event-related statistics and several useful modes of visualisation of the averaged and single-trial data.

**pyMVPA**
*http://www.pymvpa.org/*
pyMVPA (Hanke et al., 2009) is a Python package intended to ease statistical learning analyses of large datasets. It offers an extensible framework with a high-level interface to a broad range of algorithms for classification, regression, feature selection, data import and export. While it has been most popular for analysing fMRI neuroimaging datasets, it has also been used with EEG, MEG and extracellular recordings.

### eConnectome
*http://econnectome.umn.edu/*

eConnectome (He et al., 2011) is a MATLAB package for imaging brain functional connectivity from electrophysiological signals. It provides interactive graphical interfaces for EEG/ECoG/MEG pre-processing, source estimation, connectivity analysis and visualisation.

### EP_den
*http://www.vis.caltech.edu/~rodri/EP_den/EP_den_home.htm*

EP_den (Quian Quiroga and Garcia, 2003)(Quian Quiroga and Garcia, 2003) is a MATLAB package for denoising single trial evoked potentials from the background EEG. It is based on a wavelet multiresolution decomposition. It provides an interactive graphical interface for denosing the evoked potentials and visualizing the single trial responses.

### Synchro
*http://www.vis.caltech.edu/~rodri/Syncœhro/Synchro_home.htm*

Synchro (Quian Quiroga et al., 2002) is a MATLAB package for evaluating linear and non-linear synchronization measures between two signals. The nonlinear synchronization measures are based on a phase space reconstruction of the signal, which is sensitive to non-linear interactions and can in principle disclose driver-response relationships. It provides an interactive graphical interface for visualizing the data and synchronization.

## Software for Analysis of fMRI

Analysis of neuroimaging data is a huge field and a comprehensive list of software and techniques in this area is beyond the scope of this brief neural coding directory. However, for completeness we highlight below three major packages commonly used to analyse imaging data.

### SPM
*http://www.fil.ion.ucl.ac.uk/spm/*

SPM (Friston, 2007) is a MATLAB package for analysis of brain imaging data sequences, either from different cohorts or time-series from the same subject. It is designed for the analysis of fMRI, PET, SPECT, EEG and MEG.

### FSL
*http://www.fmrib.ox.ac.uk/fsl/*

FSL (Smith et al., 2004) is a comprehensive library of analysis tools for fMRI, MRI, and DTI brain imaging data. FSL is a framework of stand-alone command-line analysis programs and graphical data viewing tools and runs on UNIX-like systems.

### NIPy
*http://nipy.sourceforge.net/*

NIPy (Millman and Brett, 2007) is a Python project for analysis of structural and functional neuroimaging data.